\documentclass[a4paper]{llncs}

\usepackage{amssymb}
\usepackage[vlined,linesnumbered]{algorithm2e}
\usepackage[pdftex]{graphicx}
\usepackage{lmodern}
\usepackage{url}
\usepackage{comment}
\usepackage{subfigure}
\usepackage{paralist}

\urldef{\mailsa}\path|{williamv, sgarg, raj}@csse.unimelb.edu.au|

\newcommand{\keywords}[1]{\par\addvspace\baselineskip
\noindent\keywordname\enspace\ignorespaces#1}

\begin{document}

\mainmatter

\title{Provisioning Spot Market Cloud Resources to Create Cost-effective Virtual Clusters}

\author{William Voorsluys\inst{1} \and Saurabh Kumar Garg\inst{1} \and Rajkumar Buyya\inst{1}}

\institute{Cloud Computing and Distributed Systems (CLOUDS) Laboratory\\Department of Computer Science and Software Engineering\\The University of Melbourne, Australia\\
\mailsa\\
\url{http://www.cloudbus.org}
}

\maketitle

\begin{abstract}
Infrastructure-as-a-Service providers are offering their unused resources in the form of variable-priced virtual machines (VMs), known as ``spot instances'', at prices significantly lower than their standard fixed-priced resources. To lease spot instances, users specify a maximum price they are willing to pay per hour and VMs will run only when the current price is lower than the user's bid. This paper proposes a resource allocation policy that addresses the problem of running deadline-constrained compute-intensive jobs on a pool of composed solely of spot instances, while exploiting variations in price and performance to run applications in a fast and economical way. Our policy relies on job runtime estimations to decide what are the best types of VMs to run each job and when jobs should run. Several estimation methods are evaluated and compared, using trace-based simulations, which take real price variation traces obtained from Amazon Web Services as input, as well as an application trace from the Parallel Workload Archive. Results demonstrate the effectiveness of running computational jobs on spot instances, at a fraction (up to 60\% lower) of the price that would normally cost on fixed priced resources.
\keywords{spot market, virtual clusters, cloud computing}
\end{abstract}

\section{Introduction}
\label{intro}

The introduction of affordable cloud computing infrastructure has had a major impact in the business IT community. These resources are also being explored as a means of accomplishing high performance processing tasks, often present in areas such as science and finance. However, the use of virtualization and network shaping have been cited as factors that hinder the viability of these resources for running compute intensive applications, as opposed to using a dedicated HPC cluster \cite{hill2009}. Nonetheless, the potential cost savings offered by clouds has led to an increased adoption of cloud-based virtual clusters, as well as to the practice of extending the capacity of local clusters using cloud resources in situations of peak demand \cite{assuncao2009}.

The cost to build these virtual clusters highly depends on the type of leased virtual machines, which are offered via various pricing schemes through separate ``markets'', most notably: the on-demand market, which offers standard VMs at a fixed cost; and the spot market, which offers unused capacity in the form of variable price VMs. For example, Amazon EC2 offers biddable virtual machines (VMs), also known as ``spot instances'', for as low as $\frac{1}{3}$ of the standard fixed price for a similar instance. In this fashion, users submit a request that specifies a maximum price (bid) they are willing to pay per hour and instances associated to that request will run for as long as the current spot price is lower than the specified bid. Prices vary frequently, based on supply and demand.

In spite of the apparent economical advantage, an intermittent nature is inherent to biddable resources, which may cause VM unavailability. When an out-of-bid situation occurs, i.e. the current spot price goes above the user's maximum bid, spot instances are terminated by the provider without prior notice. However, this situation can be avoided by bidding slightly higher, thus mitigating this uncertainty, or by using fault-tolerance techniques such as checkpointing \cite{yi2010reducing}.

\begin{figure}[!hb]
\centering
\includegraphics[scale=0.5]{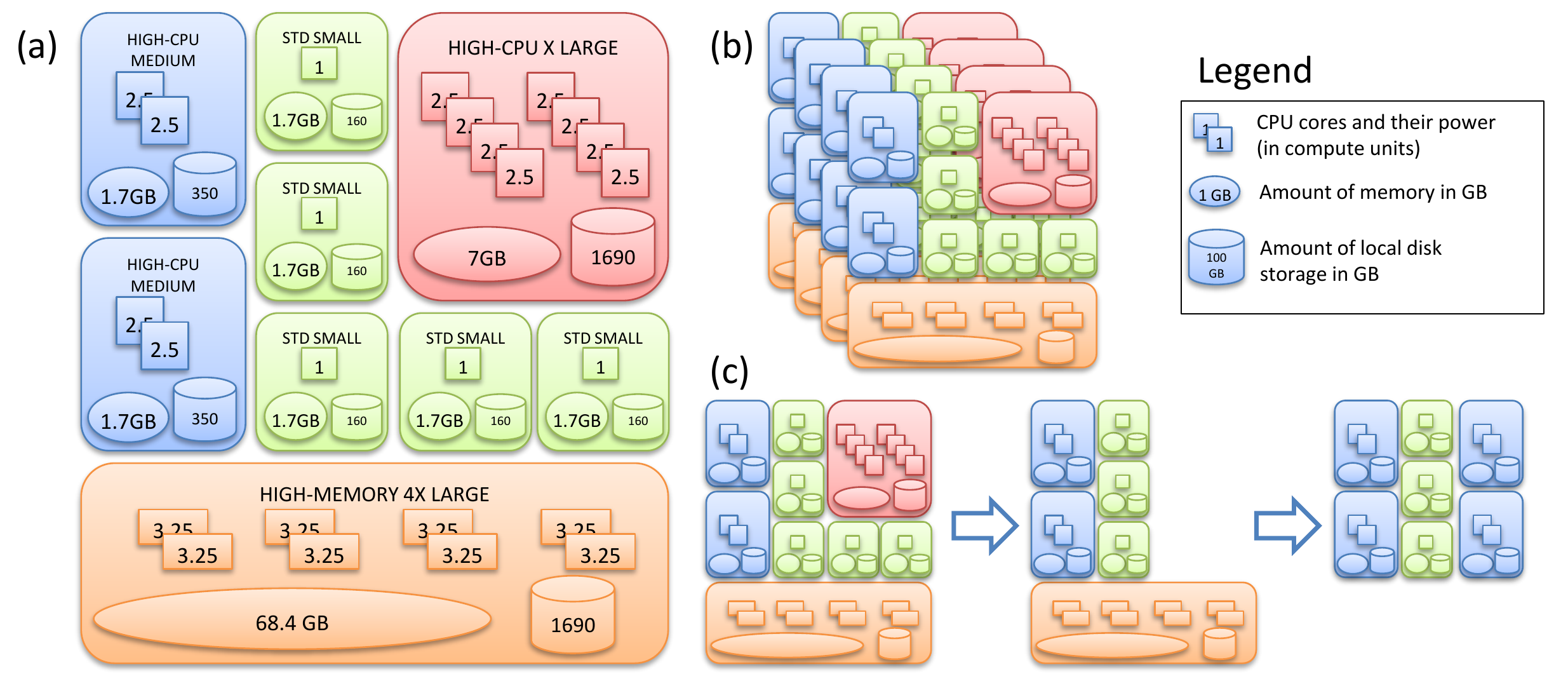}
\caption{A virtual cluster is dynamically assembled out of inexpensive cloud-based resources, e.g. Amazon EC2 spot instances. (a) VMs of appropriate sizes have to be chosen, depending on the immediate requirements of running applications; (b) the cluster is able to scale to much larger capacities; (c) VMs are replaced by different types as application requirements change}
\label{fig:scen}
\end{figure}

Virtual clusters can be heterogeneous, when different types of VMs (e.g. with distinct number of CPU cores) are leased and added to the resource pool. In this case, the ratio between price and performance of different types of spot instances may not be constant over time, creating opportunities for optimizations. For example, one could decide how to run a certain compute intensive job by observing the performance per dollar ratio of two high-CPU EC2 spot instances. The ``c1.medium'' instance type has a CPU power of 5 ECUs and the ``c1.xlarge'' type has a power of 20 ECUs. One ECU is defined as equivalent to the power of a 1.0-1.2 GHz 2007 AMD Opteron or 2007 Intel Xeon processor. As the spot price of each instance type varies, the performance per dollar ratio offered by each instance type varies accordingly so that, at different periods of time, one instance type may offer a better ratio than the other. In these situations, if an application that would normally run using 5 ECUs could provide enough parallelism, it could run significantly faster on the relatively cheaper 20 ECU instance. This approach offers extra flexibility to users since they may choose to assemble a pool of VMs by bidding on resource types that are currently at a discounted hourly price and then adapt their jobs to run more efficiently on the new resources. Figure \ref{fig:scen} depicts an example scenario of such a virtual cluster composed of virtual machines of different sizes.

This paper focuses on reducing the costs of building virtual clusters by leasing spot market resources. Specifically, we propose a \textbf{resource provisioning and job scheduling strategy that addresses the problem of dynamically building a virtual cluster out of low-cost VMs and utilizing them to run compute-intensive applications}. Our main objective is to exploit variations in price and performance of resources to run applications in a fast and economical way. Moreover, applications are assumed to be deadline-constrained. For this reason, our strategy is augmented by a job runtime estimation mechanism that aids in deciding how to allocate jobs in a way that they finish within their deadlines.

Our results show that it is possible to run a stream of computational jobs by solely utilizing spot instances. Especially, we have quantified the effect of the accuracy of runtime estimation on the monetary cost. We have found out that less accurate estimations usually lead to higher costs (in the case of over estimations) or more missed deadlines (in the case of under estimations).

\textbf{The contributions of this work are:}
\begin{itemize}
\item A novel system architecture that enables the creation of cloud-based virtual clusters solely by utilizing low-cost spot instances. Our system allows organizations that do not have a local cluster to run streams of computational jobs in a fast and economical way;
\item A resource provisioning strategy that decides when to request spot instances to accommodate incoming jobs, as well as which instance type to request.
\item An information mechanism that aids decision-making by providing runtime estimates;
\end{itemize}

The rest of this paper is organized as follows: Section \ref{sec:related} describes related literature; Section \ref{sec:system_model} describes the proposed architecture; Section \ref{sec:mech} details the mechanisms that composed our resource allocation policy; Section \ref{sec:eval} presents experimental results and their discussion; finally Section \ref{sec:conclusion} concludes the paper.

\section{Related Work}
\label{sec:related}

With the increasing popularity of cloud infrastructure, many organizations have started looking into the exploitation of cloud resources rather than maintaining their own in-house cluster facilities, which are expensive to maintain. In this section we present the most relevant works that consider similar scenarios and compared them to our contribution.

\subsection{Cloud-based virtual clusters}
Research on building virtual clusters using cloud resources can be generally divided into two categories: (1) techniques to extend the capacity of in-house clusters at times of the peak demand, and (2) assembling resource pools using only public cloud resources and using them to run compute intensive applications. For instance, Assuncao et al \cite{assuncao2009} have evaluated a set of well known scheduling policies, including backfilling techniques, in a system that extends the capacity of a local cluster using fixed-priced cloud resources. Similarly, Mattess et al \cite{mattess2010} have evaluated policies that offload extra demand from a local cluster to a resource pool composed by Amazon EC2 spot instances. In contrast to these works, our system model does not consider the existence of a local cluster, instead all resources are cloud-based spot instances.

\subsection{Use of variable pricing resources}
A few recently published works have touched the subject of leveraging variable pricing cloud resources in high-performance computing.
Andrzejak et al.~\cite{Andrzejak2010} have proposed a probabilistic decision model to help users decide how much to bid for a certain spot instance type in order to meet a given monetary budget or a deadline. The model suggests bid values based on the probability of failures calculated using a mean of past prices from Amazon EC2. It can then estimate, with a given confidence, values for a budget and a deadline that can be achieved if the given bid is used. 

Yi et al.~\cite{yi2010reducing} proposed a method to reduce costs of computations and providing fault-tolerance when using EC2 spot instances. Based on the price history, they simulated how several checkpointing policies would perform when faced with out-of-bid situations. Their evaluation has shown that checkpointing schemes, in spite of the inherent overhead, can tolerate instance failures while reducing the price paid, as compared to normal on-demand instances.

\section{System Model of a Cloud-based Virtual Cluster}
\label{sec:system_model}

We consider a scenario where an organization is interested in building a dynamic cluster using cloud resources. We have assumed that resources are to be leased exclusively from one cloud provider, such as Amazon EC2, even though our proposed solution can be extended to support multiple providers.

Resources are virtual machines, instantiated according to a previously configured template, which defines the capacity required (or ``instance type'' in Amazon EC2 terminology) as well as the operating system and software stack details (i.e. Amazon Machine Image). A job scheduling and middleware system, called Broker, is responsible for receiving job requests from users, creating a suitable VM pool, and managing the QoS of jobs, i.e. ensuring that jobs finish within the deadline.

Jobs are submitted by local users of the organization. A job request must contain information such as: the task(s) to be executed (e.g. binary files or scripts), the number of required processors, the amount of memory needed, total runtime estimation, and a deadline.

\begin{figure*}[b]
\centering
\includegraphics[scale=0.5]{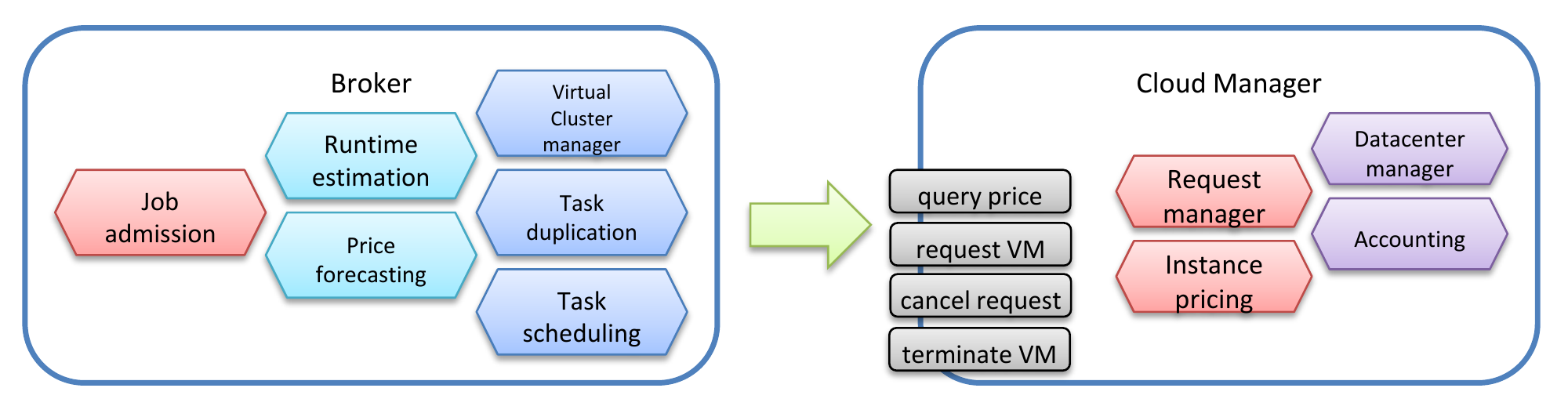}
\caption{Modeled architecture: Client (broker) and server (cloud) side components}
\label{fig:comp}
\end{figure*}

The system model we define in this work has two main components: a Broker (job scheduling and middleware); and a cloud provider-side VM management infrastructure that we term as the ``cloud manager''. A graphical representation of the modeled system components is presented on Figure \ref{fig:comp}.

\subsubsection*{Broker component:}
Computational job execution is managed by the Broker, which takes the responsibility of receiving job submissions and assembling a pool of cloud spot instances on behalf of the organization. The broker obtains all available information about a job and uses that information to perform scheduling decisions.

More specifically, the broker manages the following activities:
\begin{inparaenum}[i)]
\item Job management: job admission, job execution, job failures, and monitoring of job QoS constraints;
\item Virtual cluster management: bidding, instance selection, instance requesting and termination,
\end{inparaenum}

\subsubsection*{Cloud Provider Model:}
Spot instance management activities are performed by the cloud provider. For this component, we assume a similar cloud model as the one described by Yi et al.~\cite{yi2010reducing}, which reflects how the spot market currently works in the Amazon Cloud. More specifically, the model considers the following characteristics:

\begin{itemize}
\item Clients submit a spot instance request, specifying how many instances are needed, an instance type, and up to how much they are willing to pay per instance/hour (bid);
\item The system provides spot instances whenever the client's bid is greater than the currently advertised price; on the other hand, it terminates instances that do not meet the current spot price, immediately without any notice, when a client's bid is less than or equal to the current price.
\item The system does not charge the last partial hour when it stops an instance, but it charges the last partial hour when the termination is initiated by the client (the price of a partial hour is considered the same as a full hour). The price of each instance/hour is the spot price at the beginning of the hour.
\end{itemize}

In summary, the cloud manager is responsible for the following tasks:
\begin{inparaenum}[i)]
\item Request management: admission control based on bids, and serving valid requests;
\item Spot instance management: new instance requests, terminations due to out-of-bid situations, and billing of hourly charges.
\end{inparaenum}

\section{Cost-effective Resource Provisioning and Scheduling Policy}
\label{sec:mech}

The Broker is equipped with a VM provisioning and job scheduling policy, which composes the core contribution of this work. The following steps are involved in allocating each incoming job to a virtual machine. These activities are described in Algorithm \ref{algo:main}.

\scriptsize
\begin{algorithm}
\label{algo:main}
  \SetKwFunction{FindFreeSpace}{FindFreeSpace}
  \SetKwFunction{ComputeLeaseExtensions}{ComputeLeaseExtensions}
  \SetKwFunction{AllocateJobToVM}{AllocateJobToVM}
  \SetKwFunction{LeaseNewVM}{LeaseNewVM}
  \SetKwFunction{ComputeCostForANew}{ComputeCostForANew}
  \SetKwFunction{CanPostpone}{CanPostpone}
  \SetKwInOut{Input}{input}\SetKwInOut{Output}{output}
  \BlankLine
  \Input{available instance types $types$}
  \While{true} {
    \While{$currentTime < NEXT\_SCHEDULE\_TIME$} {
      Receives incoming jobs and inserts in the list $LJ$
    }
    
    $vms \leftarrow$ all VMs currently in the pool\;
    
    \ForEach{Job $j \in LJ$} 
    {
      $erts \leftarrow$ compute estimated runtime of $j$ on each $type \in types$\;     
      $decision \leftarrow$ \FindFreeSpace{$j$, $vms$}\;
      \If{$decision.allocated$ = true } {
	\AllocateJobToVM{$j$, $decision.vm$}\;
	continue\;
      }
      $mwt \leftarrow$ compute maximum wait time for $j$\;
      \If{\CanPostpone{$j$}} {
	Delay allocation of $j$ by $mwt$\;
	Add $j$ to list $LNU$ of non urgent jobs\;
	Remove $j$ from $LJ$\;
	continue\;
      }
      $extendDecision$ = \ComputeLeaseExtensions{$j$, $vms$}\;
      $newDecision$ = \ComputeCostForANew{$j$}\;
      
      \eIf{$extendDecision.cost <= newDecision.cost$}{
        trigger lease extension\;
	\AllocateJobToVM{$j$, $extendDecision.vm$}\;
      }{
	$newVM \leftarrow$ \LeaseNewVM{}\;
	\AllocateJobToVM{$j$, $newVM$}\;
      }
    }
    
    update state of VMs\;
    \If{there are idle VMs {\bf and} $LNU$ is not empty}{
      \ForEach{Job $j \in LNU$} {
	$decision \leftarrow$ \FindFreeSpace{$j$, $idlevms$}\;
	\If {$decision.allocate = true$} {
	  \AllocateJobToVM{$j$, $decision.vm$}\;
	  remove $j$ from $LNU$\;
	  add $j$ to $LJ$
	}
      }
    }
    
     \ForEach{Job $j \in LJ$} {
	\If {$j.isAllocate = true$} {
	  dispatch correction and rescheduling event at time $now + j.ert$
	}
     }
     
     clear $LJ$
    
    NEXT\_SCHEDULE\_TIME = NEXT\_SCHEDULE\_TIME + t\;
  }
\caption{Resource provisioning and job scheduling algorithm}
\end{algorithm}

\normalsize

\begin{itemize}
\item When a job is submitted, it is inserted into a list of unscheduled jobs [Line 3].
\item At each scheduling interval, the algorithm uses a runtime estimation method to predict the runtime of the job on each available instance type [Line 6].
\item The broker then attempts to allocate the job to an idle, already initiated, VM with enough time before a whole hour finishes [Lines 7-10].
\item If unsuccessful, it decides whether the allocation decision can be postponed, based the maximum time the job can wait so that the chance of meeting the deadline is increased [Lines 11-16].
\item If the job cannot be postponed, it attempts to allocate it to a VM that is is expected to become idle soon. Runtime estimates of all jobs running on the VM are required at this step [Lines 18-25];
\item If the job still cannot be allocated, the algorithm will decide whether to extend a current lease or to start a new one. This decision is made based on information about the estimated runtime of the incoming job on each VM type [Lines 18-25].
\item The algorithm then checks if there are still idle VMs, which are then matched to non-urgent jobs that have been postponed in previous iterations [Lines 28-34]. Idle VMs are the ones that have been flagged to be terminated when the next whole hours finishes.
\item Finally, for each job that was allocated to a VM in the current iteration, a correction event is scheduled to trigger at the time the job is expected to finish [Line 35-37]. This event, if necessary, will then activate the estimation correction and rescheduling mechanism, which corrects the estimation and reinserts all jobs allocated to the affected instance into the list of unscheduled jobs.
\end{itemize}

These steps are conducted in regular intervals (time $t$, which can be defined based on the arrival rate of jobs). In this work, we have set $t$ to 10 seconds, so that our algorithm operates in on-line fashion, especially to guarantee that jobs with a tight deadline are given to opportunity to start as soon as possible. Still, the use of techniques such as job postponing, runtime estimation and delayed termination of idle instances ensures that the policy keeps a holistic view of the workload.

Details of the various steps performed by our algorithm, such as runtime estimation and correction, rescheduling, and job speedup characteristics are given in following sections.

\subsection{Estimating Job Runtimes}

In order to circumvent inaccurate user-supplied estimations, especially harmful in backfilling schedulers, several works have proposed methods to predict job runtimes, where the system computes the estimated runtime of a job and uses it in place of a user-supplied estimate. Although  complex methods have been proposed, Tsafir~\cite{tsafrir2010} has observed that ``even extremely trivial algorithms (e.g. using the average runtime of two preceding jobs by the same user) result in significant improvement''.

Although we do not make use of backfilling techniques in this work, our motivation to employ runtime estimates is similar to those schedulers: improve system utilization, especially important because there is a minimum cost involved for each new VM added to the pool. Due to an hourly billing fashion, assumed in this work to reflect practices of cloud computing providers, every VM runs for a minimum of one hour. Therefore, relying solely on user provided runtimes (mostly over-estimated) might lead to unnecessary requests.

In our scenarios, runtime predictions aid the decision-making process in the following ways:
\begin{inparaenum}[i)]
\item the broker can decide whether to add new resources to the pool based on the expected time that currently busy VMs would be free;
\item along with information about current instance prices, it is possible to estimate the cost to run a job on a given instance type, thus increasing the chances of meeting monetary constraints.
\end{inparaenum}

Previous studies have shown that the ``one size fits all'' notion does not apply to runtime estimation of job runtimes. For this reason, we have implemented 5 different methods, especially because no method has been shown to work well in all scenarios~\cite{tsafrir2010}. A detailed discussion of each of these methods is presented in the evaluation section of this paper.

\subsection{Estimation correction and rescheduling}
\label{sec:resched}
We have equipped the broker with a correction and rescheduling mechanism, which activates whenever a job is detected to have been running longer than expected, regardless of which runtime estimation method has provided the estimate. Whenever a job starts running, a correction event is scheduled to trigger immediately after the moment the job should have finished. If, at that moment, the job is still running, a correction operation is performed. The broker simply assumes a new estimate that is equal to double the old estimate, and the job remains allocated to the current VM. All other affected jobs, i.e. the ones scheduled to the same instance, which might be delayed, are resubmitted to the scheduler for rescheduling.

\subsection{Job moldability and speedup considerations}

We assume jobs to be moldable, i.e. they can execute on any number of processing units, but restricted to a single VM. We model instance types as to contain one or more processing units, assumed to be equal to the amount of EC2 compute units of each available instance type. Each job runs on a single instance, and each instance can only run one job at a time.

To determine the runtime of a job in a particular number of compute units, we use Downey's analytical model for job speedup~\cite{downey1997}. Downey's model requires two parameters: the average parallelism A and an approximation to the coefficient of variance of parallelism \textit{$\sigma$}. To generate values for A and $\sigma$, we have used the model of Cirne \& Berman~\cite{cirne2001}, which has been shown to capture the behavior of a range of user jobs. Generated values were added as parameters to each job originally present in the LCG workload trace. We assume that these values are known by the users who submit the jobs, thus they can be used by the resource provisioning strategies.

\section{Performance Evaluation}
\label{sec:eval}

In this section, we evaluate the proposed resource allocation and scheduling policy using trace-driven discrete event simulation which is implemented using the CloudSim framework~\cite{buyya2009modeling}. The overall objective of our experiments is to quantify the performance of our proposed policy based on three metrics: monetary cost, system utilization, and deadline misses. Since our proposed policy aims at minimizing the cost of building virtual clusters, the monetary cost of such activity is considered as the main metric. System utilization indicates how long instances remain idle before they are terminated. Deadline misses refers to the number of submitted jobs which did not finish within the specified deadline; this is a metric directly related to user satisfaction.

In a first scenario, the policy is compared with two base provisioning policies: worst-case and best-case resource provisioning. In the worst-case provisioning, the broker provisions only on-demand fixed-price instances but schedules jobs on the most efficient machines types for each job. This provisioning is very similar to the current solutions for building virtual cluster using cloud resources~\cite{mattess2010}. The best-case resource provisioning is a hypothetical lower bound devised to evaluate the cost-effectiveness of the proposed policy.

In a second experimental scenario, the effects of various runtime estimation on the proposed policy are evaluated to understand which runtime estimation method should be used for a given workload.

\subsubsection*{Virtual machine types:}
As stated earlier, our resource provisioning strategy aims at choosing the most efficient instance type to run a job. Five instance types were used in our experiments. They were modeled directly after the characteristics of available standard and high-CPU Amazon EC2 types. The types available to be used are M1.SMALL (1 ECU), M1.LARGE (5 ECUs), M1.XLARGE (8 ECUs), C1.MEDIUM (5 ECUs), C1.XLARGE (20 ECUs).

\subsubsection*{Workload:} 
The chosen job stream was obtained from the LHC Grid at CERN~\cite{feitelson71parallel}, and is composed of grid-like embarrassingly parallel tasks. A total of 100,000 jobs were submitted over a period of seven days of simulation time. This workload is suitable to our experiments to due to its bursty nature and for being composed of highly variable job lengths. These features require a highly dynamic computation platform that must serve variable loads while maintaining cost efficiency.
Originally, this workload trace did not contain information about user-supplied job runtime estimates and deadlines. User runtime estimates were generated according to the model of Tzafrir et al.~\cite{Tsafrir05modelinguser}. A job's maximum allowed runtime corresponds to the runtime estimate multiplied by a random multiplier, uniformly generated between 1.5 and 4. Consequently, the job deadline corresponds to its submission time plus its maximum allowed runtime.

\subsection{Comparison with Best-case and Worst-case scenarios}

In this section, we compare our proposed scheduling policy with other base policies. Based on information from the workload trace (actual job length and parallelism parameters A and $\sigma$), we have computed how much would be the best possible cost that could be achieved to run all 100,000 jobs using the most efficient instance type for each job, considering multiple pricing schemes. The most efficient match for a job depends on its maximum speedup, the job's length and the instance's cost per hour. As a general rule, short jobs or jobs that provide little or no parallelism run more efficiently on less powerful, cheaper instances; whereas longer jobs (execution time in the order of hours) that provide good parallelism are more suitable for high-CPU instances, which provide a lower cost per ECU.

Table \ref{tab:baseline} lists the costs that would be obtained in both best-case and worst-case resource provisioning scenarios. Particularly, the cost of \$2790.28 corresponds to the best possible. Therefore, the aim of any resource provisioning strategy is to obtain a cost as close as possible to this value.

Our proposed policy, when running with the \textit{``Recent Average''} estimation method was able to obtain an improvement of about 60\% over the worst case provisioning policy and just 23\% worse than the best case. 

\begin{table}
\centering
\caption{Total cost compared with two base policies}
\label{tab:baseline}
\scriptsize
\begin{tabular}{|l| r| r| r| r|}
		\hline
		Instance type & Percentage of jobs & Worst-case  & Best-case & Proposed Policy \\
		\hline
		M1.SMALL (1 ECU) & 6.646\% & \$1114.62 & \$371.54 &NA\footnotemark\\
		C1.MEDIUM (5 ECUs) & 84.564\% & \$6942.38 & \$2314.13&NA \\
		C1.XLARGE (20 ECUs) & 8.790\% & \$313.84 & \$104.61&NA \\
		\hline
		& \textbf{Total:} & \$8370.84 & \$2790.28 & \$3628.25\\
		\hline
\end{tabular}
\end{table}

\footnotetext{We report individual percentage of jobs that ran on each instance type only for the deterministic scenarios (worst-case and best-case) as an indication of the bias towards high-cpu instances. These values are not necessarily meaningful of how the policy allocates jobs in practice, where the total cost is the metric that really matters.}

\subsection{Effect of different runtime estimation methods}

We now describe in detail the 5 runtime estimation schemes and compare their effects on the following metrics: monetary cost, number of deadline misses, and system utilization. All values presented correspond to an average of 30 simulation runs. Each run is set to start at a random point in time, uniformly chosen between March 1st, 2010 and February 1st, 2011. These dates correspond to the available price traces obtained from Amazon EC2.

In the \textit{``Actual runtime"} approach, the actual job length, as per the workload trace, is supplied to the allocation algorithm; while the \textit{``Actual runtime with error"} approach consists of using the actual length slightly modified by a random percentage between 0 and 10\%. Naturally, these two approaches are not real as they are based on information that would normally not be available in practice. They are included here for comparison purposes. However, should a nearly perfect estimation method be available, say in a highly controlled environment where detailed information about the workload characteristics is known, we can then foresee that our proposed allocation algorithm would perform well, as these two strategies yield the best results.

The \textit{``User Supplied"} approach assumes the job length to be the value informed by the user at job submission. Based on previous observations that user-supplied estimated runtimes are mostly over estimated, we have also devised the \textit{``Fraction of User Supplied"} approach, that uses a value equal to $\frac{1}{3}$ of original value as the job length.

\begin{figure}[!b]
\centering
\includegraphics[scale=0.19]{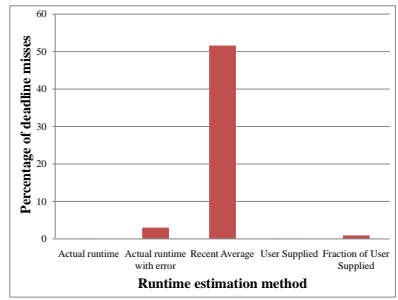}
\caption{Deadline misses before the correction and rescheduling mechanism was introduced}
\label{fig:rt_est_deadlines}
\end{figure}

The \textit{``Recent Average"} approach consists of using the average runtime of two jobs completed prior to the submission of an incoming job by the same user. If not enough information is available to compute the estimated length of an incoming job, i.e. less than two jobs have completed at the time of decision-making, the estimation is assumed to be given by the \textit{``User Supplied"} method.

We conducted two sets of experiments. In first set, our strategy was not equipped with the correction and rescheduling mechanism, described on section \ref{sec:resched}. In these experiments, once the strategy made a decision based on a runtime estimate, jobs would remain allocated to the instance chosen on the first decision. This resulted in an excessive amount of deadline misses, especially when using the \textit{``Recent Average''} estimation method, as can be see on Figure \ref{fig:rt_est_deadlines}. This fact is due to under estimations, that caused many jobs to be allocated to the same instances. Once a certain job that was expected to finish at a certain time did not finish, all other jobs would be delayed. By adding correction and rescheduling, the strategy was able to virtually eliminate the occurrence of deadline misses.

\begin{figure}[!b]
\centering
\subfigure[]{\includegraphics[scale=0.2]{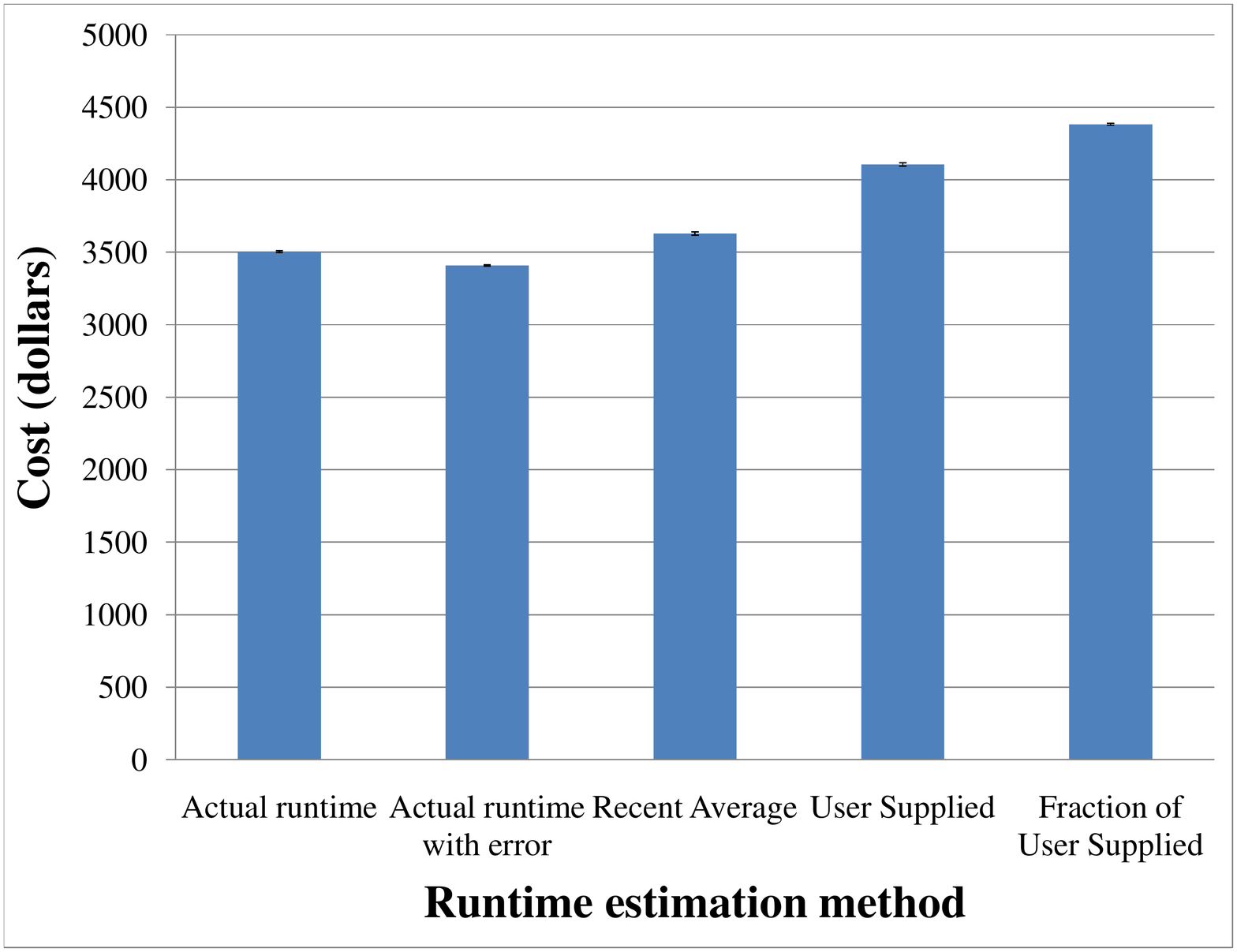}
\label{fig:rt_est_cost}}
\subfigure[]{\includegraphics[scale=0.2]{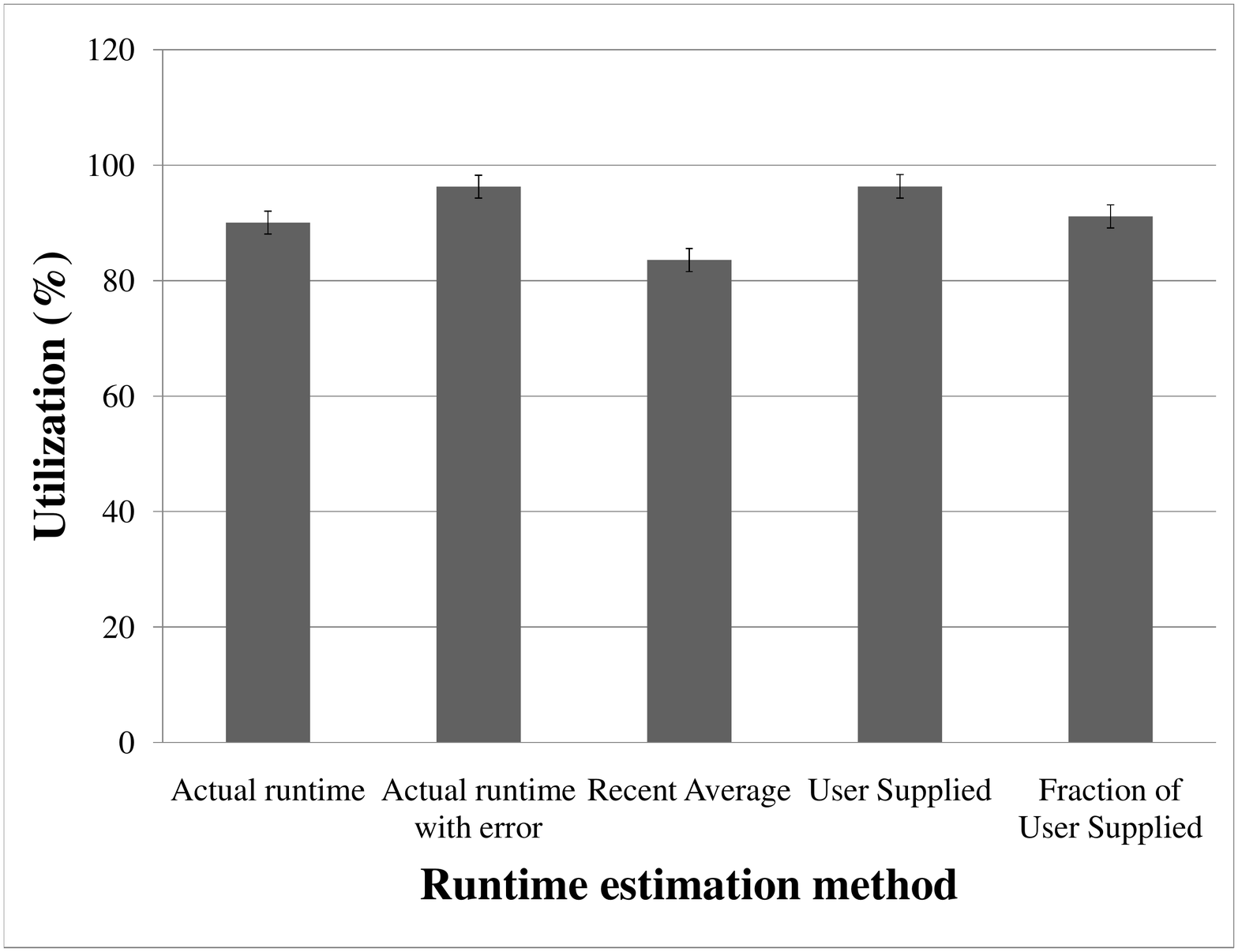}
\label{fig:rt_est_util}}
\caption{Effect of different runtime estimation methods; \subref{fig:rt_est_cost} on monetary cost, and \subref{fig:rt_est_util} on system utilization}
\label{fig:timings}
\end{figure}

Figures \ref{fig:rt_est_cost}, and \ref{fig:rt_est_util} show the effect of changing the runtime estimation component on the total monetary cost, and system utilization respectively. These results correspond our second set of experiments, which were collected after the correction and rescheduling mechanism was implemented.

Results demonstrate that, contrary to our early belief, precise information does not necessarily translates into more efficient allocation, especially in terms of cost. This fact can be observed in the comparison between the \textit{``Actual runtime"} and \textit{``Actual runtime with error"} methods, where the latter performs better.  Based on observations of the simulation logs, we can attribute this difference to moderate over-estimations, that cause more instances to be requested, which are reused often by short jobs. On the other hand, the exact estimates provided by the optimal method cause the allocation system to request fewer instances and also to terminate instances more often. Incoming jobs then cause more new instances to be requested, which are then more likely to remain idle if the job that triggered the request was a short one. This fact can be confirmed by observing the system utilization under the effects of the \textit{``Actual runtime"} method (90\%) and \textit{``Actual runtime with error"} (96\%).

In term of deadline misses, runtime estimations methods that tend to over-estimate provide better results. However, excessively over-estimated runtimes were shown to increase costs significantly, as they cause a much higher number of instances to be requested, especially more expensive instances. Although these instances are sometimes reused by other jobs, in most cases the allocation algorithm will choose to create new instances, by mistakenly considering that incoming jobs are long, thus requiring more powerful instances to complete their execution within their deadlines. This observation can be inferred from the results obtained by the \textit{``User Supplied"} and \textit{``Fraction of User Supplied"} estimation methods, also presented on \ref{fig:rt_est_cost}; both methods tend to provide excessively over-estimated runtimes.

We have also observed that good system utilization alone does not necessarily translates into lower costs. For example, by using the \textit{``Recent Average"} estimation method, our strategy could achieve a better cost than \textit{``User Supplied''}, but the utilization was significantly lower. The key aspect to observe in this scenario is the importance of choosing the correct type of instances for each job. Smaller instances, even when not utilized efficiently, have less impact on the final cost, than larger and more costly instance, which must be used efficiently to compensate for their cost.

In conclusion, in our scenarios, slightly over-estimated runtimes have shown to be beneficial. On the other hand, excessively over-estimations are ineffective because of the bias towards larger and costly instances. Under-estimations have shown to increase the chance of deadline misses, but they are not as ineffective, as they can be corrected and affected jobs can be rescheduled. In any case, we can conclude that accurate estimations help in achieving lower costs.

\section{Conclusions and Future Work}
\label{sec:conclusion}

Building dynamic virtual clusters using cloud resources is an effective way of saving in monetary cost. This paper has provided a cost-effective solution to provision a virtual cluster using spot market cloud resources to run computational jobs having a deadline as a QoS constraint. To address this challenge, a resource allocation and job scheduling policy, which takes into account variations in price and performance of cloud resources and aims at choosing the most efficient virtual machines for deadline-constrained jobs. We have evaluated the performance of the proposed policy, by comparing it to worst-case and best-case scenarios. An improvement of up to 60\% in cost has been obtained in comparison with the worst-case, and the policy has performed close to the best-case. 

We have also evaluated 5 runtime estimation methods and their effects on the policy performance. For the given workload, we have concluded that more accurate estimation methods provide significantly superior results, when compared to methods that excessively over-estimate runtimes.

As future work, we will integrate our proposed solution into real cloud schedulers, and evaluate further performance benefits of our approach. We will also tackle the problem of jobs failures and delays due to the intermittent nature of spot instances, by applying fault tolerance techniques such as workload migration, checkpointing and task duplication.

\bibliographystyle{splncs}

\bibliography{spotsim}

\begin{thebibliography}{10}

\bibitem{hill2009}
Hill, Z., Humphrey, M.:
\newblock {A quantitative analysis of high performance computing with Amazon's
  EC2 infrastructure: The death of the local cluster?}
\newblock In: Proceedings of the 10th IEEE/ACM International Conference on Grid
  Computing. (oct. 2009)

\bibitem{assuncao2009}
de~Assuncao, M.D., di~Costanzo, A., Buyya, R.:
\newblock Evaluating the cost-benefit of using cloud computing to extend the
  capacity of clusters.
\newblock In: Proceedings of the 18th ACM International Symposium on High
  Performance Distributed Computing. HPDC '09, New York, NY, USA, ACM (2009)

\bibitem{yi2010reducing}
Yi, S., Kondo, D., Andrzejak, A.:
\newblock Reducing costs of spot instances via checkpointing in the amazon
  elastic compute cloud.
\newblock In: 2010 IEEE 3rd International Conference on Cloud Computing, IEEE
  (2010)  236--243

\bibitem{mattess2010}
Mattess, M., Vecchiola, C., Buyya, R.:
\newblock {Managing Peak Loads by Leasing Cloud Infrastructure Services from a
  Spot Market}.
\newblock In: Proceedings of the 10th IEEE International Conference on High
  Performance Computing and Communications, Los Alamitos, CA, USA, IEEE
  Computer Society (2010)  180--188

\bibitem{Andrzejak2010}
Andrzejak, A., Kondo, D., Yi, S.:
\newblock {Decision Model for Cloud Computing under SLA Constraints}.
\newblock Technical report, INRIA (2010)

\bibitem{tsafrir2010}
Tsafrir, D.:
\newblock {Using inaccurate estimates accurately}.
\newblock In: Symposium on Job Scheduling Strategies for Parallel Processing,
  JSSPP 2010, Springer (2010)  208--221

\bibitem{downey1997}
Downey, A.B.:
\newblock {A Model For Speedup of Parallel Programs}.
\newblock Technical report, Berkeley, CA, USA (1997)

\bibitem{cirne2001}
Cirne, W., Berman, F.:
\newblock {A Model for Moldable Supercomputer Jobs}.
\newblock In: Proceedings of the 15th International Parallel and Distributed
  Processing Symposium, Los Alamitos, CA, USA, IEEE Computer Society (2001)

\bibitem{buyya2009modeling}
Buyya, R., Ranjan, R., Calheiros, R.:
\newblock {Modeling and simulation of scalable cloud computing environments and
  the cloudsim toolkit: Challenges and opportunities}.
\newblock In: Proceeding of the International Conference on High Performance
  Computing \& Simulation, HPCS'09, IEEE (2009)  1--11

\bibitem{feitelson71parallel}
Feitelson, D.:
\newblock Parallel workloads archive.
\newblock http://www.cs.huji.ac.il/labs/parallel/workload

\bibitem{Tsafrir05modelinguser}
Tsafrir, D., Etsion, Y., Feitelson, D.G.:
\newblock {Modeling User Runtime Estimates}.
\newblock In: In 11th Workshop on Job Scheduling Strategies for Parallel
  Processing, JSSPP, Springer-Verlag (2005)  1--35

\end{thebibliography}

\end{document}